\long\def\UN#1{$\underline{{\vphantom{\hbox{#1}}}\smash{\hbox{#1}}}$}
\def\NP{\vfil\eject}
\def\NI{\noindent}

\magnification=\magstep 2
\overfullrule=0pt
\hfuzz=16pt
\voffset=0.0 true in
\vsize=8.8 true in
\baselineskip 20pt
\parskip 6pt
\hoffset=0.1 true in
\hsize=6.3 true in
\nopagenumbers
\pageno=1
\footline={\hfil -- {\folio} -- \hfil}
 
\hphantom{A} 

\centerline{\UN{\bf Extended Quantum XOR Gate in Terms
of Two-Spin Interactions}}
 
\vskip 0.4in
 
\centerline{{\bf Dima Mozyrsky,}\ \ {\bf Vladimir Privman}}
 
\centerline{\sl Department of Physics, Clarkson University,
Potsdam, NY 13699--5820}
 
\centerline{{\bf and}}

\centerline{{\bf Steven P. Hotaling}}
 
\centerline{\sl Air Force Materiel Command, Rome
Laboratory/Photonics Division}
\centerline{\sl 25 Electronic Parkway, Rome, NY 13441--4515}
 
\vfill
 
\centerline{\bf ABSTRACT}
 
Considerations of feasibility of quantum computing lead to the study
of multispin quantum gates in which the input and output two-state
systems (spins) are not identical. We provide a general discussion of
this approach and then propose an explicit {\it two-spin interaction\/}
Hamiltonian which accomplishes the quantum XOR gate function for a
system of three spins: two input and one output.

\vfill
 
\NI {\bf PACS numbers:}\ 03.65.Bz, 89.80.+h, 02.20.Sv, 76.70.Fz
 
\NP

The size of semiconductor computer components is still quite far from
atomic dimensions. They will soon reach [1] linear dimensions of about
$0.25\,\mu$m. This is $2500\,$\AA, well above
the sizes at which quantum-mechanical effects will be important.
It has become clear, however, that as the miniaturization continues,
atomic dimensions will be reached, perhaps, with technology different
from today's semiconductors. Then, quantum-mechanical effects will have
to be considered in computer operation. 
Lead by this expectation, some early works [2-4] considered how
quantum mechanics affects the foundations of computer science.
Questions such as limitations on ``classical'' computation due to
quantum fluctuations, etc., have been raised.

A more ``active'' approach, initiated recently by many authors
[4-30] is to attempt to harness the quantum nature of components
of atomic
dimensions for more efficient computation and design. This ambitious
program involves many interesting scientific concepts new to both
Computer Science and Physics.
In order to answer whether quantum computation is feasible and useful,
several issues must be addressed. Is quantum computation faster
than classical computation? Can quantum computational elements be
built and combined with other quantum and/or classical components?
What will be the ``design'' rules for quantum computer
components in order to perform Boolean logic operations on quantum
bits (qubits) such as the up and down spin
states of a spin-$1\over 2$ particle?
What are the error correction requirements and methods in
quantum computation?

The answers to some of the questions that result from consideration of
these general issues are still in the future. However, many definitive
results have already been obtained. Specifically, on the theoretical
side, new fast quantum algorithms have been proposed [31-35]. Error
correction techniques [10,27,31,36,37], unitary operations corresponding
to the simplest logic gates [5-30], and some Hamiltonians for gate
operation [10,14,24,28-30] have
been explored. Ideas on how to combine the simplest quantum gates
have been put forth, e.g., [7,15,38]. On the side of experiment,
there are
several atomic-scale systems where the simplest quantum-gate functions
have been recently realized [26,39,40] or contemplated [19].

There are, however, many general [4,18] and specific problems both
with theory and experimental realizations of quantum computing.
Just to mention one of them, the reversibility of coherent quantum
evolution implies that the time scale $\Delta t$ of the operation of
quantum logic gates must be built into the Hamiltonian. As a result,
virtually all proposals available to date assume that computation will
be externally timed, i.e., interactions will be switched ``on''
and ``off,'' for instance, by laser radiation. 

Thus, while we deem it inevitable that quantum
properties of matter on the atomic scale will have to be considered in
computer component design and use, we recognize that it
is still a long way to
go, with modern technology, to a really ``desktop'' fully coherent
quantum computational unit. We propose to adopt a more realistic
expectation that 
technological advances will first allow design and manufacturing of
limited-size units, based on several tens of atomic two-level systems,
operating in a quantum-coherent fashion over a large time interval and
possibly driven externally by laser beams. These units will then become
parts of a larger ``classical'' computer which will not maintain a
quantum-coherent operation over its macroscopic dimensions.

A program of study should therefore begin with the simplest quantum
logic gates in order to
identify which Hamiltonians are typical for interactions required for
their operation. Results presently available are limited;
they include Hamiltonians for certain NOT [14,28] and 
controlled-NOT gates [10,30], and for some
copying processes [29,30], as well as general analyses of possibility
of construction of quantum operations [8,22]. In order to make
connections with the present ``classical'' computer-circuitry design
rules and have a natural way of identifying, at least initially,
which multi-qubit systems are of interest,
we propose to consider spatially extended quantum gates, i.e.,
gates with input and output qubits different.

Of course, reversibility of coherent quantum evolution makes the
distinction between the input and output less important than in
irreversible present-day
computer components. However, we consider this notion useful within
our general goals: to learn what kind of interactions are involved
and to consider also units that might be connected to/as in
``classical'' computer devices. While our present study is analytical,
we foresee studies of systems of order 20 to 25 two-state (qubit) atomic
``components'' with general-parameter interactions identified in the
earlier work. Then, by using ordinary computers one can design
those interaction
values for which the resulting computational units will be useful
as part of a computer and will be usable for Boolean logic operations
(this need of numerical calculations limits the number
of constituents to 20-25, i.e., to systems with total of $2^{20}$ to
$2^{25}$ states that modern ``classical'' computers can handle). 

A more futuristic goal of incorporating such
computational units in actual computer design will require a whole
new branch of computer engineering because the ``built-in'' Boolean
functions will be quite complicated as compared to the present-day
components such as NOT, AND, OR, NAND, to which computer designers
are accustomed. Furthermore, the rules of their
interconnection with each other and with the rest of the ``classical''
computer will be different from today's devices.

In this Letter we consider the XOR gate. Let us use the term ``spin''
to describe a two-state system, and we will represent spin-$1\over
2$-particle spin-components (measured in units of $\hbar /2$) by
the standard Pauli
matrices $\sigma_{x,y,z}$. In Figure 1, we denote by $A$, $B$, $C$ the
three two-state systems, i.e., three spins, involved. We assume that
at time $t$
the input spins $A$ and $B$ are in one of the basis states $|AB\rangle=
|11\rangle$, $|10\rangle$, $|01\rangle$, or $|00\rangle$, where 1 and
0 denote the eigenstates of the $z$-component of the spin operator,
with 1 referring to the ``up'' state and 0 referring to the ``down''
state. We use this notation for consistency with the classical ``bit''
notion.
The initial state of $C$ is not specified (it is arbitrary). 

We would like to have a quantum evolution which, provided $A$ and
$B$ are initially in those basis states, mimics the truth table of XOR:

$$\matrix{A&B&{\rm output}\cr{}1&1&0\cr{}1&0&1\cr{}0&1&1\cr{}0&0&0}
\eqno(1) $$

\NI were the output is at time $t+\Delta t$. One way to accomplish
this is to produce the output in $A$ or $B$, i.e.,
work with a two-spin system where the input and output are the same.
The Hamiltonian for such a system is not unique. Explicit examples
can be found in [10,30] where XOR was obtained as a sub-result of the
controlled-NOT
gate operation. In this case of two spins involved, the interactions
can be single- and two-spin only.

An important question is whether multispin systems can produce useful
logical operations with only two-spin interactions. Indeed,
two-particle interactions are much better studied and accessible to
experimental probe than multiparticle interactions. Here we report
such an example for the three-spin system depicted in Figure 1. To our
best knowledge, this is the first such result in the literature.

Thus, we require that the XOR result will be generated in $C$ at time
$t+\Delta t$. The final states of $A$ and $B$, as well as the
{\it phase\/} of $C$ are not really specified (they are arbitrary). In
fact, there are many different unitary transformations, $U$,
that correspond to the desired evolution in the eight-state space with
the basis $|ABC\rangle=|111\rangle$, $|110\rangle$, $|101\rangle$,
$|100\rangle$, $|011\rangle$, $|010\rangle$, $|001\rangle$,
$|000\rangle$, which we will use in this order. The choice of the
transformation determines what happens when the initial state is a
superposition of the reference states, what are the phases in
the output, etc. The transformation is definitely not unique.

Consider the following Hamiltonian,

$$ H={\pi \hbar \over 4 \Delta t}
 \left( \sqrt{2} \sigma_{zA} \sigma_{yB}
+ \sqrt{2} \sigma_{zB} \sigma_{yC}-\sigma_{yB} \sigma_{xC}
\right) \eqno(2)$$

\NI It is written here in terms of the spin components. In the
eight-state basis specified earlier, its matrix can be obtained by
direct product of the Pauli matrices and unit $2\times 2$ matrices
$\cal I$. Here the subscripts $A,B,C$ denote the spins. For
instance, the first interaction term is proportional to

$$ \sigma_{zA}\otimes\sigma_{yB}\otimes{\cal I}_C \eqno(3)$$

\NI etc. This Hamiltonian involves only two-spin-component interactions.
In fact, $A$ and $C$ only interact with $B$, see Figure 1, so
diagrammatically there in no loop (it is not known if the latter
property is significant since we are dealing here with
``nonequilibrium,'' i.e., non-ground-state, calculations).

One can show that the Hamiltonian (2) corresponds to the XOR result
in $C$ at $t+\Delta t$ provided $A$ and $B$ where in one of the
allowed superpositions of the appropriate ``binary'' states at $t$
(we refer to superposition here because $C$ is arbitrary at $t$).
There are two ways to verify this claim. Firstly, one can diagonalize
$H$ directly, calculate $U$ in the diagonal representation by using
the general relation (valid for Hamiltonians which are constant
during the time interval $\Delta t$; see [28] for a formulation that
introduces a multiplicative time dependence
in $H$),

$$ U=\exp \left(-iH\Delta t /\hbar \right) \eqno(4) $$

\NI and then reverse the diagonalizing transformation. The result for
$U$, as a matrix in the basis selected earlier, is

$$ U
=\pmatrix{0&0&-1&0&0&0&0&0\cr{}1&0&0&0&0&0&0&0\cr{}0&0&0&1&0&0&0&0\cr
0&1&0&0&0&0&0&0\cr{}0&0&0&0&0&-1&0&0\cr{}0&0&0&0&0&0&0&1\cr
0&0&0&0&-1&0&0&0\cr{}0&0&0&0&0&0&-1&0} \eqno(5)$$

\NI The calculation is extremely cumbersome (it was carried out, in
part, by using Maple symbolic computer language);
we do not reproduce it here.

The second, more general approach, by which the form (2) was actually
obtained originally, is to analyze generally $8\times 8$ unitary
matrices
corresponding to the XOR evolution, i.e., any liner combination of
the states $|\underline{11}1\rangle$ and
$|\underline{11}0\rangle$ evolves
into a linear combination of $|11\underline{0}\rangle$,
$|10\underline{0}\rangle$, $|01\underline{0}\rangle$, and
$|00\underline{0}\rangle$, compare the underlined quantum numbers with
the first entry in (1), with similar rules for the other three entries
in (1). One can conjecture and analyze forms that yield two-spin
interaction Hamiltonians. This approach is also quite cumbersome 
and not particularly illuminating. It will be detailed elsewhere [41];
the result is a three-parameter family of two-spin XOR Hamiltonians
[41] from which we selected (2) as a particularly elegant and
short (and also ``loopless'' in the sense mentioned earlier) form. 

It is quite straightforward to check that, with phase factors $-1$ in
some cases, the unitary matrix $U$ indeed places the XOR$(A,B)$ in
$C$. Note that (2) is not symmetric in $A$ and $B$, so that another
Hamiltonian can be obtained by relabeling.

In summary, we demonstrated by explicit example that two-spin
interaction Hamiltonians can be useful in generating standard
logical operations in systems with more than two spins. Analytical
results and general rules are difficult to come up with. It is likely
that future quantum logic gate ``design'' will involve heavy
numerical simulations of systems of several spins with trial
two-spin interactions, to determine interaction parameter values
for which they perform useful logical operations.

The work at Clarkson University has been supported in part by a US Air
Force grant, contract number F30602-96-1-0276. 
The work at Rome Laboratory
has been supported by the AFOSR Exploratory Research Program and by
the Photonics in-house Research Program.
This financial assistance is gratefully acknowledged.

\NP

\centerline{\bf REFERENCES}{\frenchspacing
 
\item{[1]} Scientific American (August 1996) article on page 33.

\item{[2]} R. Feynman, Int. J. Theor. 
Phys. {\bf 21}, 467 (1982).

\item{[3]} R. Feynman, Optics News 
{\bf 11}, 11 (1985).

\item{[4]} An instructive survey of these issues and references
to literature can be found in the article by R. Landauer, 
Philos. Trans. R. Soc. London Ser. 
A {\bf 353}, 367 (1995). 

\item{[5]} A. Barenco, 
Proc. R. Soc. Lond. A {\bf 449}, 679 (1995).

\item{[6]} A. Barenco, ``Quantum 
Physics and Computers'' (preprint).

\item{[7]} A. Barenco, C.H. Bennett, 
R. Cleve, D.P. DiVincenzo,
N. Margolus, P. Shor, T. Sleator, J.A. Smolin and H. Weinfurter,
Phys. Rev. A {\bf 52}, 3457 (1995).

\item{[8]} P. Benioff, J. Stat. Phys. {\bf 29}, 515 (1982).

\item{[9]} C.H. Bennett, 
Physics Today, October 1995, p. 24.

\item{[10]} I.L. Chung and Y. Yamamoto,
``The Persistent Qubit'' (preprint).

\item{[11]} J.I. Cirac and P. Zoller, 
Phys. Rev. Lett. {\bf 74}, 4091 (1995).

\item{[12]} D. Deutsch, Physics World, June 1992, p. 57.

\item{[13]} D. Deutsch, A. Barenco and A. Ekert, 
Proc. R. Soc. Lond. A {\bf 449}, 669 (1995).

\item{[14]} D.P. DiVincenzo, Science {\bf 270}, 255 (1995). 

\item{[15]} D.P. DiVincenzo, 
Phys. Rev. A {\bf 51}, 1015 (1995).

\item{[16]} A. Ekert, ``Quantum Computation'' (preprint). 

\item{[17]} A. Ekert and R. Jozsa, Rev. Mod. 
Phys. (to appear).

\item{[18]} S. Haroche and J.-M. Raimond, 
Physics Today, August 1996, p. 51.

\item{[19]} S.P. Hotaling, ``Radix-$R>2$ 
Quantum Computation'' (preprint).

\item{[20]} S. Lloyd, Science {\bf 261}, 1563 (1993). 

\item{[21]} N. Margolus, ``Parallel Quantum 
Computation'' (preprint).

\item{[22]} A. Peres, Phys. Rev. A {\bf 32}, 3266 (1985). 

\item{[23]} D.R. Simon, ``On the Power of 
Quantum Computation'' (preprint). 

\item{[24]} A. Steane, ``The Ion Trap 
Quantum Information Processor'' (preprint). 

\item{[25]} B. Schumacher, Phys. Rev. A {\bf 51}, 2738 (1995). 

\item{[26]} B. Schwarzschild, Physics Today, March 1996, p. 21. 

\item{[27]} W.H. Zurek, Phys. Rev. Lett. {\bf 53}, 391 (1984). 

\item{[28]} D. Mozyrsky, V. Privman and S.P. Hotaling,
``Design of Gates for Quantum Computation: the NOT Gate'' (preprint).

\item{[29]} D. Mozyrsky and V. Privman, ``Quantum Signal Splitting as
Entanglement due to Three-Spin Interactions'' (preprint).

\item{[30]} D. Mozyrsky, V. Privman and M. Hillery, ``A Hamiltonian
for Quantum Copying,'' (preprint).

\item{[31]} I.L. Chuang, R. Laflamme, 
P.W. Shor and W.H. Zurek,
Science {\bf 270}, 1633 (1995).

\item{[32]} C. D\"urr and P. H\o yer, 
``A Quantum Algorithm for Finding the Minimum'' (preprint).

\item{[33]} R.B. Griffiths and C.-S. Niu, 
``Semiclassical Fourier Transform for Quantum
Computation'' (preprint).

\item{[34]} L.K. Grover, ``A Fast Quantum 
Mechanical Algorithm for Estimating the Median'' (preprint).

\item{[35]} P.W. Shor, ``Algorithms 
for Quantum Computation: Discrete Log and 
Factoring. Extended Abstract.'' (preprint). 

\item{[36]} E. Knill and R. Laflamme,
``A Theory of Quantum 
Error-Correcting Codes'' (preprint). 

\item{[37]} W.G. Unruh, Phys. Rev. A {\bf 51}, 992 (1995). 

\item{[38]} S. Lloyd, Phys. Rev. Lett. {\bf 75}, 346 (1995)

\item{[39]} C. Monroe, D.M. Meekhof, B.E. King, 
W.M. Itano and D.J. Wineland, 
Phys. Rev. Lett. {\bf 75}, 4714 (1995).

\item{[40]} Q. Turchette, C. Hood, 
W. Lange, H. Mabushi and H.J. Kimble,
Phys. Rev. Lett. {\bf 75}, 4710 (1995). 

\item{[41]} D. Mozyrsky, V. Privman and S.P. Hotaling,
in preparation.

}

\NP

\centerline{\bf FIGURE CAPTION}

\NI Figure 1:\ \ Three two-state systems (spins) $A$, $B$, $C$, with
the pairwise spin-component interactions in (2) marked schematically
by the connecting lines. The XOR operation accomplished by (2) in
time $\Delta t$ assumes that $A$ and $B$ are the input qubits and
$C$ is the output qubit.
 
\bye